\documentclass[aip,jcp,preprint]{revtex4-2}
\usepackage{amsmath}            	
\usepackage{amssymb}				
\usepackage{amsmath}
\usepackage{threeparttable}
\usepackage{booktabs}
\usepackage[version=4]{mhchem}		
\usepackage{xcolor}					
\usepackage{tcolorbox}              
\usepackage[title]{appendix}
\usepackage{enumerate}          	
\usepackage{mathtools}
\usepackage{multirow}				
\usepackage{hyperref}

\newcommand{\red}[1]{\textcolor{black}{#1}}
\allowdisplaybreaks
\begin{document}
\title{Unitary coupled-cluster based self-consistent polarization propagator theory: a quadratic unitary coupled-cluster singles and doubles scheme}
\author{Junzi Liu}
\email{latrix1247@gmail.com}
\author{Lan Cheng}
\email{lcheng24@jhu.edu}
\affiliation{Department of Chemistry, The Johns Hopkins University, Baltimore, MD, 21218, USA}
\begin{abstract}

The development of a quadratic unitary coupled-cluster singles and doubles (qUCCSD) based self-consistent polarization propagator method is reported.
We present a simple strategy for truncating the commutator expansion 
of the UCC transformed Hamiltonian $\bar{H}$. 
The qUCCSD method for the electronic ground-state 
includes up to double commutators for the amplitude equations and up to cubic commutators for the energy expression.  
The qUCCSD excited-state eigenvalue equations include up to double commutators for the singles-singles block of $\bar{H}$, single commutators for the singles-doubles and doubles-singles blocks, and the bare Hamiltonian for the doubles-doubles block.
Benchmark qUCCSD calculations of the ground-state properties and excitation energies for representative molecules demonstrate significant improvement of the accuracy and robustness over the previous UCC3 scheme derived using M{\o}ller-Plesset perturbation theory.
\end{abstract}

\maketitle
\section{Introduction}

The equation-of-motion coupled-cluster (EOM-CC) methods
\cite{Emrich1981,Stanton1993,Piecuch2000,Kucharski2001,Kowalski2004,Hirata2004a,Levchenko2004,Kallay2004,Krylov2008} 
and the closely-related CC linear response (CC-LR) theory \cite{Monkhorst1977,Nakatsuji1978_JCP,Mukherjee1979,Monkhorst1983,Koch1990_CCRF,Koch1990_CCSDLR,Christiansen1995,Christiansen1998,Kallay2004} have been established as useful tools for 
treating electronically excited states of small and medium-sized molecules.
Recent efforts have also been devoted to extending the applicability of EOM-CC
and similarity-transformed EOM-CC methods\cite{Nooijen1997,Nooijen1997a}
to large molecules\cite{Helmich2013,Dutta2016,Dutta2018a,Frank2018,Izsak2020} and
solids.\cite{Hirata2008,Katagiri2005,Wang2020,Pulkin2020,Gallo2021} In spite of the tremendous
success, the non-hermitian nature of the CC theory poses difficult unsolved problems. 
CC calculations in combination with complex Hamiltonians, 
e.g., the Hamiltonian in magnetic fields and/or including spin-orbit coupling
have been shown to produce complex ground-state energies.\cite{Thomas2021complex}
This is a non-trivial formal problem of the standard CC theory, although the real part
of the complex CC energy is expected to serve as an accurate approximation
to the full configuration interaction energy.  
Further, EOM-CC calculations have been demonstrated to 
have incorrect crossing conditions for 
intersections between electronic states of the same symmetry
(known as ``same-symmetry conical intersection'').
\cite{Hattig2005,Kohn2007,Kjonstad2017-a,Thomas2021complex}
To enable CC calculations of same-symmetry conical intersections, K\"{o}hn and Tajti \cite{Kohn2007} 
have proposed a simple 
correction to obtain physically meaningful potential energy surfaces
around conical intersections.
Koch and collaborators have recently developed a similarity constrained
coupled-cluster singles and doubles (SCCSD) method that introduces an additional parameter associated with a triple excitation and determines this parameter by requiring the eigenvectors of two target states to be orthogonal to each other. 
\cite{Kjonstad2017-b,Kjostad2019} 
These correction schemes have to introduce a substantial modification 
to the \red{wavefunctions} in order to enforce orthogonalization in two
otherwise parrallel eigenvectors.
For example, the resulting SCC wavefunction often involves
a significant contribution from a triply excited determinant.\cite{Kjonstad2017-b,Kjostad2019} 
\red{On the other hand, SCCSD produces excitation energies similar to that of
CCSD.}


Same-symmetry conical intersections play essential roles in photochemistry.
\cite{Domcke2004,Levine2007,Domcke2012}
Available calculations of same symmetry conical intersections
have used hermitian excited-state formulations such as the algebraic diagrammatic construction (ADC) methods,
\cite{Hattig2005,Lefrancois2017}  
time-dependent density functional theory (TDDFT)-based techniques,\cite{Levine2006,Ou2013} 
constrained density functional theory-configuration interaction (CDFT-CI) method,\cite{Kaduk2010} 
and multireference techniques including complete active space
self-consistent-field (CASSCF) method,\cite{Ben-Nun2000} CAS second-order
perturbation theory (CASPT2),\cite{Levine2008} multi-reference perturbation
theory (MRPT),\cite{Nangia2006,Xu2014} and multi-reference configuration
interaction (MRCI) methods.\cite{Lengsfield1984,Woywod1994,Yarkony2001,Matsika2004,Zhu2014}
The MRCI method as a non-perturbative wavefunction based approach 
has exhibited robust performance. However, the lack of size-extensivity in MRCI
often poses difficulties in obtaining accurate electronic energies.
For example, while MRCI calculations provided high-quality potential energy surfaces 
to gain insights into nonadiabatic tunneling dynamics of 
phenol dissociation, \cite{Zhu2016-a,Zhu2016-b,Xie2016}
an energetic shift had to be applied to the computed potential energy surfaces
to obtain a good agreement with the experimental energetics.\cite{Zhu2016-b} 
The size-inextensivity problem of MRCI is expected to be more serious for 
calculations of larger molecules.
Therefore, 
the development of new non-perturbative size-extensive/size-intensive hermitian excited-state theories
to enhance the capability to treat same-symmetry conical intersections is of significant interest to photochemistry applications.

The unitary version of coupled-cluster (UCC) theory appears to be a natural approach to solve 
the formal problems of the CC theory arising from nonhermiticity and to enable
CC studies of same-symmetry conical intersections. 
Analyses of the formal properties for the UCC theory and the relation with the standard CC methods 
have been reported. 
\cite{Kutzelnigg1983,Kutzelnigg1991,Bartlett1989,Harsha2018,Evangelista2019} 
Numerical studies of the UCC methods have been carried out. \cite{Cooper2010,Evangelista2011}
The UCC methods truncated up to a given rank of excitation operators 
have been shown to recover a similar amount 
of dynamic correlation energies compared with 
the standard CC methods involving the same ranks of excitation
operators. \cite{Evangelista2011}
However, a formidable challenge in the UCC theory
is to develop a practically tractable truncation scheme for the non-terminating
expansion of the transformed Hamiltonian while 
maintaining the computational accuracy. 
Several truncation schemes for the ground-state UCC
theory have been reported. The UCC(4) and UCC(5) methods have been developed using
a perturbative analysis of
the UCC energy expression.\cite{Bartlett1989,Watts1989}
Taube and Bartlett have reported a truncation scheme exact
for two-electron systems.\cite{Taube2006} 
The commutator truncation schemes have been explored for the multireference
version of UCC theory \cite{Hoffmann1988, Chen2012} and the canonical transformation
methods.\cite{Neuscamman2009,Neuscamman2010}
A stochastic approach to select excitation operators in UCC calculations has recently been developed. \cite{Filip2020}
The recent development of density-cumulant functional theory has also provided information about the accuracy of the 
truncation schemes for hermitian formulations.\cite{Sokolov2013,Misiewicz2020}
We mention the rapidly growing interest in using UCC in quantum computations and refer the readers to recent publications
 and the references therein for this exploding field.
 \cite{Peruzzo2014,Wecker2015,Barkoutsos2018,Romero2019,Grimsley2019,Lee2019,
Evangelista2019,Sokolov2020,Lang2021,Chen2021,Pavosevic2021,Bauman2021}
Here accurate and efficient UCC calculations on classical computers have the potential to help the initial state preparation for quantum computations. 


Concerning UCC-based excited-state theories, 
the second-order version UCC linear response theory 
has been shown to be identical
to the second-order version of ADC [ADC(2)].\cite{Walz2012} 
We have recently developed a third-order formulation 
for calculations of both ground-state energies and excitation energies
within the UCC-based polarization propagator (PP)
framework (the UCC3 scheme).\cite{Liu2018} 
Interestingly, the strict version of UCC3 (UCC3-s) has been shown to be equivalent to
the strict version of the third order ADC [ADC(3)], \cite{Schirmer2002,Dreuw2015,Banerjee2021} establishing
the relation between the UCC-based polarization propagator theory
and ADC.\cite{Liu2018}
Hodecker {\it{et al.}} have reported an implementation 
of UCC3 \cite{Hodecker2020} and a combination with a second-order density matrix for calculations of properties. \cite{Hodecker2020a} 
Although the schemes based on perturbation theory performs well
for simple molecules around the equilibrium structures,
the performance decays quickly for more complex molecules
in the absence of smooth convergence of the M{\o}ller-Plesset (MP) perturbation series.
Therefore, we base our present work on an alternative strategy of truncating
the expansion of the UCC transformed Hamiltonian to up to a certain power of cluster amplitudes.
In Section II, We report the formulation and implementation of a quadratic UCCSD scheme (qUCCSD) 
for calculations of ground-state energies and excitation energies. 
The details about the benchmark calculations are discussed in
Section III. The benchmark results for ground-state properties and 
excitation energies are presented and discussed in section
IV. Finally, a summary and a perspective about future work are
presented in Section V.

\section{Theory}


\subsection{Unitary coupled-cluster based polarization propagator theory}


In this subsection, we present a succinct summary of unitary coupled-cluster based polarization propagator (UCC-PP) theory
in the language of wavefunction theory.
We refer the readers to Refs.~\citenum{Liu2018} 
and \citenum{Prasad1985} for a
detailed account of the UCC-PP theory and to the literature
\cite{Nooijen1992,Nooijen1993,Nooijen1995,Meissner1993,Kowalski2014,Peng2016,
Mcclain2016,Lange2018,Mertins1996,Banerjee2019}
for Green's function methods based on the biorthogonal CC representation.
The self-consistent polarization propagator methods represent the polarization
propagator in an approximate many-electron basis by applying the inner-projection technique \cite{Lowdin1971} with a
self-consistent operator manifold to decouple the forward and backward
polarization propagator.\cite{Prasad1985}
In the UCC-based self-consistent polarization propagator method,
the ground-state wavefunction adopts the UCC parameterization
\begin{equation}
	|\Psi_\mathrm{gr} \rangle = e^\sigma | \Phi_0 \rangle,
\end{equation}
in which the cluster operator $\sigma$
comprises both excitation and de-excitation operators, e.g.,
in the UCC singles and doubles (UCCSD) method $\sigma$ can be written as
\begin{eqnarray}
	\sigma &=& \sigma_1 + \sigma_2, \\
	\sigma_1 &=& \sum_{ai} \sigma_i^a \{ a_a^\dagger a_i\} - \sum_{ai}(\sigma_i^a)^\ast
	\{a_i^\dagger a_a \}, \\
	\sigma_2 &=& \frac{1}{4}\left[\sum_{abij} \sigma_{ij}^{ab} \{ a_a^\dagger a_b^\dagger a_j a_i\} 
	- \sum_{abij} (\sigma_{ij}^{ab})^\ast \{ a_a a_b a_j^\dagger a_i^\dagger \}
	\right].
\end{eqnarray}
$\{i,j,\dots\}$ and $\{a,b,\dots\}$ denote occupied orbitals and
virtual orbitals, respectively. $\sigma_i^a$ and $\sigma_{ij}^{ab}$ represent the cluster amplitudes. 
This anti-hermitian form of the cluster operator $\sigma$
ensures the wave operator $e^\sigma$ to be unitary.
The UCCSD ground-state energy and amplitude equations are given by 
\begin{eqnarray}
	\langle \Phi_0 | \bar{H} | \Phi_0 \rangle &=& E_\mathrm{gr} , \label{en_eq} \\
	\langle \Phi_l | \bar{H} | \Phi_0 \rangle &=& 0 . \label{ap_eq}
\end{eqnarray}
Here the transformed Hamiltonian $\bar{H}=e^{-\sigma} H e^\sigma$ is hermitian. 
$\Phi_0$ represents the ground-state Hartree-Fock wavefunction,
while $\Phi_l$'s denote singly and doubly excited determinants. 
The UCC-based polarization propagator theory employs a self-consistent operator manifold
consisting of the transformed excitation and de-excitation operators,
$\{e^\sigma b_I^\dagger e^{-\sigma}\} \cup \{e^\sigma b_I e^{-\sigma}\}$,
%
in which $b_I^\dagger$ is the original excitation operators, i.e., 
$\{b_I^\dagger\}=\{a_a^\dagger a_i\}\cup\{a_a^\dagger a_b^\dagger a_j a_i\}$ in
the UCCSD method. 
This leads to the following eigenvalue equations 
\begin{equation}
	\sum_I \bar{H}_{JI} C_{IL} = E_L C_{JL}~,~ \bar{H}_{JI}=\langle \Phi_0|b_J \bar{H} b_I^\dag|\Phi_0\rangle, 
	\label{secular_eq}
\end{equation}
to determine excitation energies $E_L$ and
the excited-state wavefunctions 
\begin{equation}
	|\Phi_L^{ext} \rangle =  \sum_I C_{IL} e^\sigma b_I^\dagger |\Phi_0\rangle.
\end{equation}
\red{In other words, the UCCSD excited-state equations solve for eigenvalues and
eigenstate of $\bar{H}$ within the space of singly and doubly excited determinants.}
The excited-state secular equations can be rewritten in a block form as 
\begin{eqnarray} 
  \begin{bmatrix} 
	\bar{H}_\mathrm{SS} &\bar{H}_\mathrm{SD} \\ 
	\bar{H}_\mathrm{DS} &\bar{H}_\mathrm{DD}
  \end{bmatrix} 
  \begin{bmatrix} 
	  C_\mathrm{S} \\ C_\mathrm{D} 
  \end{bmatrix} = E
  \begin{bmatrix} 
	  C_\mathrm{S} \\ C_\mathrm{D} 
  \end{bmatrix}. \label{eigen_eq}
\end{eqnarray}
Here $\bar{H}_\mathrm{SS}$ refers to the singles-singles block  
involving $\bar{H}_{ij}$, $\bar{H}_{ab}$, and $\bar{H}_{ia,bj}$,
$\bar{H}_\mathrm{SD}$ and $\bar{H}_\mathrm{DS}$ represent the singles-doubles block and
doubles-singles block involving the contributions from $\bar{H}_{ci,ab}$, $\bar{H}_{jk,ia}$,
$\bar{H}_{ajk,ibc}$ and $\bar{H}_{ab,ci}$, $\bar{H}_{ia,jk}$, $\bar{H}_{ibc,ajk}$,
and $\bar{H}_\mathrm{DD}$ is the doubles-doubles block involving 
$\bar{H}_{ij}$, $\bar{H}_{ab}$, $\bar{H}_{ia,bj}$, $\bar{H}_{ij,kl}$,
$\bar{H}_{ab,cd}$, $\bar{H}_{iab,jcd}$, and
$\bar{H}_{ija,klb}$. 
 The $\bar{H}$ components pertinent to 
the UCCSD ground-state energy and amplitude equations as well 
as the excited-state secular equations thus can be summarized as 
\begin{eqnarray}
	\bar{H} 
	&=& E_\mathrm{gr}+ \left((\bar{H}_{ai}\{a_a^\dag a_i\}
	+\frac{1}{4} \bar{H}_{ab,ij}\{a_a^\dag a_b^\dag a_j a_i\})+h.c.\right) \nonumber \\
	&+& 
	\left(\bar{H}_{ij}\{a_i^\dag a_j\} + \bar{H}_{ab}\{a_a^\dag a_b\}
	+\frac{1}{4} \bar{H}_{ij,kl}\{a_i^\dag a_j^\dag a_l a_k\}
	+\frac{1}{4} \bar{H}_{ab,cd}\{a_a^\dag a_b^\dag a_d a_c\}
	+            \bar{H}_{ia,bj}\{a_i^\dag a_a^\dag a_j a_b\} \right) \nonumber \\
	&+& 
	\left( 
	(\frac{1}{2} \bar{H}_{ij,ka}\{a_i^\dag a_j^\dag a_a a_k\}
	+\frac{1}{2} \bar{H}_{ab,ci}\{a_a^\dag a_b^\dag a_i a_c\}
	)+h.c.\right) \nonumber \\
	&+& 
	\left( \frac{1}{4} \bar{H}_{ibc,ajk}\{a_i^\dag a_b^\dag a_c^\dag a_k a_j a_a \}+h.c.\right) \nonumber \\
	&+&\left( \frac{1}{4} \bar{H}_{iab,jcd}\{a_i^\dag a_a^\dag a_b^\dag a_d a_c a_j \} 
	+\frac{1}{4} \bar{H}_{ija,klb}\{a_i^\dag a_j^\dag a_a^\dag a_b a_l a_k \} \right), 
	\label{hbarexp} 
\end{eqnarray}
in which $E_{gr}$ is the UCCSD ground-state energy. 

In contrast to that $\bar{H}$ in the CC theory 
terminates at the quadruple commutators, the commutator expansion of $\bar{H}$ in the UCC theory
is non-terminating. 
We adopt an expansion using Bernoulli numbers 
for $\bar{H}$ \cite{Liu2018}
\begin{eqnarray}
	\bar{H} &=& \bar{H}^0 + \bar{H}^1 + \bar{H}^2 + \bar{H}^3 + \bar{H}^4 + \cdots\cdots, \\
	\bar{H}^0 &=& F + V, \\
	\bar{H}^1 &=& [F, \sigma] + \frac{1}{2}[V, \sigma] + \frac{1}{2}[V_R, \sigma], \\
	\bar{H}^2 &=& \frac{1}{12}[[V_N, \sigma], \sigma] + \frac{1}{4}[[V, \sigma]_R, \sigma] 
				+ \frac{1}{4}[[V_R, \sigma]_R, \sigma], \\
	\bar{H}^3 &=& \frac{1}{24}[[[V_N, \sigma], \sigma]_R,\sigma] + \frac{1}{8}[[[V_R, \sigma]_R, 
	              \sigma]_R, \sigma] + \frac{1}{8}[[V, \sigma]_R, \sigma]_R, \sigma] \nonumber \\
			  & & - \frac{1}{24}[[[V, \sigma]_R, \sigma],\sigma]
				  - \frac{1}{24}[[[V_R, \sigma]_R, \sigma],\sigma], \\
	\bar{H}^4 &=& \frac{1}{16}[[[[V_R, \sigma]_R, \sigma]_R,\sigma]_R,\sigma] + 
	              \frac{1}{16}[[[[V,   \sigma]_R, \sigma]_R,\sigma]_R,\sigma] + 
	              \frac{1}{48}[[[[V_N, \sigma]  , \sigma]_R,\sigma]_R,\sigma] \nonumber \\ 
	          & & -\frac{1}{48}[[[[V,  \sigma]_R, \sigma],  \sigma]_R, \sigma] 
			      -\frac{1}{48}[[[[V_R,\sigma]_R, \sigma],  \sigma]_R, \sigma]
				  -\frac{1}{144}[[[[V_N, \sigma], \sigma]_R,\sigma], \sigma]  \nonumber  \\
	          & & -\frac{1}{48}[[[[V,  \sigma]_R, \sigma]_R,\sigma], \sigma] 
			      -\frac{1}{48}[[[[V_R,\sigma]_R, \sigma]_R,\sigma], \sigma]
				  -\frac{1}{720}[[[[V_N, \sigma], \sigma] ,\sigma], \sigma].
\end{eqnarray}
Here ``N'' refers to the joint set of excitation and de-excitation portions of the target operator, while ``R'' refers to the rest of the operator excluding the ``N'' part. \cite{Liu2018} This expansion using Bernoulli numbers eliminates higher than linear commutators with respect to the Fock operator and offers a compact framework for formulating practical UCCSD methods.

\subsection{A general strategy for truncating the commutator expansion and the qUCCSD scheme}

The magnitude of the cluster amplitudes serves as a faithful measure for the strength of dynamic correlation. We thus explore UCC truncation schemes based on the powers of the cluster amplitudes, or equivalently, on the order of commutators in the commutator expansion of $\bar{H}$ using Bernoulli numbers. 
Note that, although $\sigma_1$ emerges at the second order in M{\o}ller-Plesset
perturbation theory, single-reference systems with strong orbital-relaxation
effects exhibit large ground-state CC amplitudes for single excitations. The
standard CC methods can provide accurate treatments of orbital relaxation
through the exponential of single excitations. \cite{Thouless1960} However,
methods based on MP perturbation theory or truncation of single excitations to
the linear terms could not treat these systems accurately, e.g., see Refs.
\cite{Bohme1994,Hrusak1997} Therefore, we truncate single and double excitations
up to the same power in the present work. We use a general notation
UCCSD[$k$$\mid$$l$,$m$,$n$] to denote a scheme that include up to the $k$'th order
commutators for the ground-state amplitude equations [($k$+1)'th order commutators
for the ground-state energy expression], $l$'th order commutators for the
singles-singles block of the excited-state secular equations, $m$'th order for the
singles-doubles and doubles-singles blocks, and $n$'th order commutators for the
doubles-doubles block.

Applying the partitioning technique\cite{Lowdin1971} to Eq.~\eqref{eigen_eq}
to fold the contributions from double excitations into singly excited states,
the eigenvalue equations can be rewritten as 
\begin{equation}
	\left(\bar{H}_\mathrm{SS} + \bar{H}_\mathrm{SD}(E-\bar{H}_\mathrm{DD})^{-1}
	\bar{H}_\mathrm{DS}\right)C_\mathrm{S} = EC_\mathrm{S}.
\end{equation}
\red{A balanced truncation scheme for the excited-state eigenvalue equations
thus would involve expansions of $\bar{H}_\mathrm{SS}$ and 
$\bar{H}_\mathrm{SD}(E-\bar{H}_\mathrm{DD})^{-1}\bar{H}_\mathrm{DS}$ to the same
accuracy. Since $V$ serves as a similar measure of electron correlation as
$\sigma$, we count the power of $V$ and $\sigma$ together. $\bar{H}_\mathrm{SS}$
and $\bar{H}_\mathrm{DD}$ involve $F$, $V$, and commutators of $V$ and
$\sigma$. The expansions of $\bar{H}_\mathrm{SS}$ and
$(E-\bar{H}_\mathrm{DD})^{-1}$ thus start with a contribution of $F$, which is
of the zeroth power of $V$ and $\sigma$. In contrast, $\bar{H}_\mathrm{SD}$ and
$\bar{H}_\mathrm{DS}$ involve $V$ and commutators of $V$ and $\sigma$, and thus 
are of at least linear power.} 
\red{Therefore}, the truncation of $\bar{H}_\mathrm{SS}$ to the $l$'th order commutators
\red{of $V$ and $\sigma$}, $\bar{H}_\mathrm{SD}$ and $\bar{H}_\mathrm{DS}$ to the 
($l$-1)'th order commutators, and $\bar{H}_\mathrm{DD}$ to the ($l$-2)'th order commutators ensures
$\bar{H}_\mathrm{SS}$ and $\bar{H}_\mathrm{SD}(E-\bar{H}_\mathrm{DD})^{-1}\bar{H}_\mathrm{DS}$
to be correct up to the \red{$l+1$}'th power of \red{$V$} and $\sigma$ and 
provides a balanced description for the singly excited states. 
Further, \red{we choose to include in $\bar{H}_\mathrm{SS}$ the same ranks of
commutators} as in the ground-state amplitude equations.
The UCCSD[$l$$\mid$$l$,$l$-1,$l$-2] schemes thus emerge as \red{promising} options
for treating ground state and singly excited states.
Since the linearized methods usually are numerically not accurate, in the present work we explore
the quadratic version, UCCSD[2$\mid$2,1,0], which we will refer to
as the qUCCSD scheme.

We should mention that the present general strategy is also applicable 
to the Baker-Campbell-Hausdorff (BCH) expansion.
Since $[F,\sigma]$ is of similar magnitude as $V$, 
the commutators between $F$ and $\sigma$ in the BCH expansion
should be truncated to one rank higher than the commutators between $V$
and $\sigma$. 
For example, the qUCCSD scheme within the BCH expansion consists of
up to quadruple commutators for $F$ and $\sigma$ for the energy expression,
triple commutators of $F$ and $\sigma$ for the ground-state amplitude equations,
triple commutators of $F$ and $\sigma$ for $\bar{H}_\mathrm{SS}$,
double commutators of $F$ and $\sigma$ for $\bar{H}_\mathrm{SD}$ and $\bar{H}_\mathrm{DS}$,
and single commutators of $F$ and $\sigma$ for $\bar{H}_\mathrm{DD}$\red{.}
The expansion using Bernoulli numbers is more compact than
the BCH expansion. On the other hand,
the BCH expansion is applicable to non-Hartree-Fock reference functions.
The present work is focused on the qUCCSD scheme with the expansion
using Bernoulli numbers.

\subsection{The working equations for the qUCCSD scheme}

\red{The qUCCSD working equations have been derived using the recipe for
$\bar{H}$ discussed in the previous subsection and the standard diagrammatic
techniques as in the CC methods.\cite{Crawford2000,Shavitt2009}}
The expression for the qUCCSD ground-state energy $E_{\text{gr}}^{\text{qUCCSD}}$ consists of
up to the third commutators of the fully contracted part
of $\bar{H}$ and can be written as
\begin{eqnarray}
	E_{\text{gr}}^{\text{qUCCSD}} &=&  E_\mathrm{HF} 
	          + \langle \Phi_0|\bar{H}^{1}|\Phi_0\rangle
	          + \langle \Phi_0|\bar{H}^{2}|\Phi_0\rangle
	          + \langle \Phi_0|\bar{H}^{3}|\Phi_0\rangle, \\
    \langle \Phi_0|\bar{H}^{1}|\Phi_0\rangle
    &=&\sum_{ijab} \frac{1}{8}  \langle ij||ab \rangle 
				  \sigma_{ij}^{ab} + h.c., \label{en_1}  \\
    \langle \Phi_0|\bar{H}^{2}|\Phi_0\rangle&=&\sum_{ijab} \frac{1}{12} \langle ij||ab \rangle 
				  \sigma_{i}^{a} \sigma_{j}^{b} + h.c., \label{en_2}\\
    \langle \Phi_0|\bar{H}^{3}|\Phi_0\rangle&=&\Bigg(
\Big( - \sum_{ijklabcd} \frac{1}{12} (\sigma_{kl}^{cd})^* \langle ij||ab \rangle 
        \sigma_{ik}^{ac} \sigma_{jl}^{bd}
     {+}\sum_{ijklabcd} \frac{1}{24} (\sigma_{kl}^{cd})^* \langle ij||ab \rangle 
	    \sigma_{ij}^{ac} \sigma_{kl}^{bd} \nonumber \\
&&\phantom{+\Bigg(}
   {+}\sum_{ijklabcd} \frac{1}{24} (\sigma_{kl}^{cd})^* \langle ij||ab \rangle 
	  \sigma_{ik}^{ab} \sigma_{jl}^{cd}
    - \sum_{ijklabcd} \frac{1}{96} (\sigma_{kl}^{cd})^* \langle ij||ab \rangle 
	  \sigma_{ij}^{cd} \sigma_{kl}^{ab} \Big)
	  + h.c.\Bigg) \nonumber \\ 
&+&\Bigg(\Big( \phantom{-}
      \sum_{ijklabc} \frac{1}{4} (\sigma_{jl}^{bc})^* \langle ij||ak \rangle 
      \sigma_i^a\sigma_{kl}^{bc}
	- \sum_{ijkabcd} \frac{1}{4} (\sigma_{jk}^{cd})^* \langle ic||ab \rangle 
	  \sigma_i^a\sigma_{jk}^{bd}     \nonumber \\
&&\phantom{+\Bigg(}
    + \sum_{ijklabc} \frac{1}{2}  (\sigma_{il}^{bc})^* \langle kj||ai \rangle 
	  \sigma_{j}^{b}\sigma_{lk}^{ca}
    - \sum_{ijkabcd} \frac{1}{2}  (\sigma_{jk}^{cd})^* \langle ic||ab \rangle 
	  \sigma_{j}^{b}\sigma_{ik}^{ad} \nonumber \\
&&\phantom{+\Bigg(}
    + \sum_{ijklabc} \frac{1}{12} (\sigma_{il}^{bc})^* \langle jk||ia \rangle \sigma_{l}^{c}\sigma_{kj}^{ab}
    - \sum_{ijkabcd} \frac{1}{12} (\sigma_{jk}^{cd})^* \langle ic||ab \rangle \sigma_{k}^{d}\sigma_{ij}^{ab} \nonumber \\
&&\phantom{+\Bigg(}
    - \sum_{ijklabc} \frac{1}{8}  (\sigma_{il}^{cb})^* \langle kj||ia \rangle \sigma_{l}^{a}\sigma_{kj}^{cb}
    + \sum_{ijkabcd} \frac{1}{8}  (\sigma_{jk}^{dc})^* \langle ci||ab \rangle
	\sigma_{i}^{d}\sigma_{kj}^{ab} \Big) + h.c.\Bigg)
	\nonumber \\
&+&\Bigg(\Big(
	- \sum_{ijkabc} \frac{1}{12} (\sigma_{k}^{c})^* \langle ij||ab \rangle \sigma_i^a\sigma_{jk}^{bc}
	+ \sum_{ijkabc} \frac{1}{12} (\sigma_{k}^{c})^* \langle ij||ab \rangle \sigma_i^c\sigma_{jk}^{ba}     \nonumber \\
&&\phantom{+\Bigg(}
    + \sum_{ijkabc} \frac{1}{12} (\sigma_{k}^{c})^* \langle ij||ab \rangle \sigma_{k}^{a}\sigma_{ij}^{cb}
    + \sum_{ijkabc} \frac{1}{3}  (\sigma_{ik}^{bc})^* \langle jb||ai \rangle \sigma_{j}^{c}\sigma_{k}^{a} \nonumber \\
&&\phantom{+\Bigg(}
    - \sum_{ijklab} \frac{1}{12} (\sigma_{ij}^{ab})^* \langle kl||ij \rangle \sigma_{k}^{a}\sigma_{l}^{b}
    - \sum_{ijabcd} \frac{1}{12} (\sigma_{ij}^{cd})^* \langle cd||ab \rangle
	\sigma_{j}^{b}\sigma_{i}^{a} \Big)  + h.c.\Bigg)
	\nonumber \\
&+&\Bigg(\Big(
    + \sum_{ijkab}  \frac{1}{3}  (\sigma_{k}^{b})^* \langle ij||ak \rangle \sigma_{i}^{a}\sigma_{j}^{b}
    - \sum_{ijabc}  \frac{1}{3}  (\sigma_{j}^{a})^* \langle ai||bc \rangle
	\sigma_{j}^{b}\sigma_{i}^{c} \Big)  + h.c.\Bigg). \label{en_3_7th}
\end{eqnarray}
The qUCCSD amplitude equations 
comprise up to double commutators for $\bar{H}_{ai}$
\begin{eqnarray} 
    &&\bar{H}^{\text{qUCCSD}}_{ai}=\bar{H}_{ai}^{1}+\bar{H}_{ai}^{2}=0, \\
    \bar{H}_{ai}^{1} &=&
	\red{\sum_b f_{ab} t_i^b - \sum_j t_j^a f_{ji}}+
	\frac{1}{2}\sum_{jbc}\langle aj||cb\rangle \sigma_{ij}^{cb}-
	\frac{1}{2}\sum_{jkb}\langle kj||ib\rangle \sigma_{jk}^{ba}+
	\sum_{jb}\langle aj||ib\rangle \sigma_j^{b}+
	\frac{1}{2}\sum_{jb}(\sigma_j^b)^*\langle ab||ij\rangle, \label{hbarai1} \\
	\bar{H}_{ai}^{2} &=& - \sum_{jklbc} \frac{1}{2} (\sigma_{jk}^{bc})^* \langle
	al||ik\rangle  \sigma_{jl}^{bc} + \sum_{jkbcd} \frac{1}{2}
	(\sigma_{jk}^{bd})^* \langle ad||ic\rangle  \sigma_{jk}^{bc} - \sum_{jklbc}
	(\sigma_{jk}^{bc})^* \langle bl||ji\rangle  \sigma_{kl}^{ca} + \sum_{jkbcd}
	(\sigma_{jk}^{bc})^* \langle ab||dj\rangle  \sigma_{ki}^{cd} \nonumber \\
	&-& \sum_{jklbc} \frac{1}{4} (\sigma_{jk}^{bc})^* \langle bl||jk\rangle
	\sigma_{il}^{ac} + \sum_{jkbcd} \frac{1}{4} (\sigma_{jk}^{bd})^* \langle
	bd||jc\rangle  \sigma_{ik}^{ac} + \sum_{jklbc} \frac{1}{4}
	(\sigma_{jk}^{bd})^* \langle bd||ic\rangle  \sigma_{jk}^{ca} - \sum_{jkbcd}
	\frac{1}{4} (\sigma_{jk}^{bc})^* \langle al||jk\rangle  \sigma_{il}^{cb}
	\nonumber \\ &+& \sum_{jkbc} \frac{5}{12} \langle jk||bc\rangle \sigma_j^b
	\sigma_{ik}^{ac} - \sum_{jkbc} \frac{1}{3}  \langle jk||bc\rangle \sigma_k^a
	\sigma_{ij}^{cb} - \sum_{jkbc} \frac{1}{3}  \langle jk||bc\rangle \sigma_i^c
	\sigma_{jk}^{ba} - \sum_{jkbc} \frac{1}{2}  (\sigma_k^c)^* \langle
	cj||ib\rangle \sigma_{jk}^{ba} \nonumber \\ &-& \sum_{jkbc} \frac{1}{2}
	(\sigma_k^c)^* \langle aj||kb\rangle \sigma_{ij}^{cb} - \sum_{jkbc}
	\frac{1}{3}  (\sigma_{jk}^{cb})^* \langle ab||ij\rangle \sigma_k^c -
	\sum_{jkbc} \frac{1}{6}  (\sigma_{jk}^{bc})^* \langle bc||ji\rangle
	\sigma_k^a  - \sum_{jkbc} \frac{1}{6}  (\sigma_{jk}^{bc})^* \langle
	ab||kj\rangle \sigma_i^c \nonumber \\ &+& \sum_{jbcd} \frac{1}{4}
	(\sigma_j^c)^* \langle ac || bd \rangle \sigma_{ij}^{bd} + \sum_{jklb}
	\frac{1}{4}  (\sigma_k^b)^* \langle jl || ik \rangle \sigma_{jl}^{ab}
	\nonumber \\ &+& \sum_{jbc}            \langle aj||cb\rangle \sigma_j^{b}
	\sigma_i^c - \sum_{jkb}            \langle kj||ib\rangle \sigma_j^{b}
	\sigma_k^a + \sum_{jbc}\frac{1}{2} (\sigma_j^b)^* \langle ab||cj\rangle
	\sigma_i^c - \sum_{jkb}\frac{1}{2} (\sigma_j^b)^* \langle kb||ij\rangle
	\sigma_k^a \nonumber \\ &+& \sum_{jbc}\frac{1}{2} (\sigma_j^c)^* \langle
	ac||ib\rangle \sigma_j^b  - \sum_{jkb}\frac{1}{2} (\sigma_j^b)^* \langle
	ak||ij\rangle \sigma_k^b, \label{hbarai2} 
\end{eqnarray}
and for $\bar{H}_{ab,ij}$
\begin{eqnarray}
\bar{H}^{\text{qUCCSD}}_{ab,ij}&=&\bar{H}_{ab,ij}^0+\bar{H}_{ab,ij}^1+\bar{H}_{ab,ij}^2=0, \\
	\bar{H}_{ab,ij}^{0}&=&\langle ab || ij \rangle, \label{hbarabij0} \\
	\bar{H}_{ab,ij}^{1}
	&=&
	 \sum_c f_{ac} \sigma_{ij}^{cb} - \sum_k f_{ki} \sigma_{kj}^{ab}+
	 \frac{1}{2} \sum_{kl} \langle kl||ij \rangle \sigma_{kl}^{ab}+
	 \frac{1}{2} \sum_{cd} \langle ab||cd \rangle \sigma_{ij}^{cd}+ 
	 P(ij) P(ab) \sum_{kc} \langle ak||ic \rangle \sigma_{jk}^{bc} \nonumber \\
	&-&
	 P(ab) \sum_{k} \langle ka||ji\rangle \sigma_k^{b} +
	 P(ij) \sum_{c} \langle ab||ic\rangle \sigma_j^{c}, \label{hbarabij1} \\
	\bar{H}_{ab,ij}^{2}
	&=&
	P(ij) P(ab) \sum_{klcd} \frac{1}{3} \langle kl||cd\rangle \sigma_{ik}^{ac} \sigma_{jl}^{bd} +
	\sum_{klcd}  \frac{1}{6}  \langle kl||cd \rangle \sigma_{ij}^{cd} \sigma_{kl}^{ab} -
	P(ab) \sum_{klcd} \frac{1}{3} \langle kl||cd \rangle \sigma_{ij}^{ad} \sigma_{kl}^{cb}
	\nonumber \\
	&-&
	P(ij) \sum_{klcd} \frac{1}{3} \langle kl||cd \rangle \sigma_{il}^{ab} \sigma_{jk}^{dc} +
	P(ij) P(ab) \sum_{klcd}  \frac{1}{3} (\sigma_{kl}^{cd})^* \langle ad||il\rangle \sigma_{jk}^{bc} +
	\sum_{klcd} \frac{1}{12} (\sigma_{kl}^{cd})^* \langle cd||ij\rangle \sigma_{kl}^{ab}
	\nonumber \\
	&+&
	\sum_{klcd}  \frac{1}{12} (\sigma_{kl}^{cd})^* \langle ab||kl\rangle \sigma_{ij}^{cd} -
	P(ab) \sum_{klcd} \frac{1}{6} (\sigma_{kl}^{cd})^* \langle ad||ij\rangle  \sigma_{kl}^{cb} -
	P(ij) \sum_{klcd} \frac{1}{6} (\sigma_{kl}^{cd})^* \langle ab||il\rangle  \sigma_{jk}^{dc}
	\nonumber \\
	&-&
	P(ab) \sum_{klcd} \frac{1}{6} (\sigma_{kl}^{cd})^* \langle cb||kl\rangle \sigma_{ij}^{ad} -
	P(ij) \sum_{klcd} \frac{1}{6} (\sigma_{kl}^{cd})^* \langle cd||kj\rangle \sigma_{il}^{ab}
	\nonumber \\
	&-&
	P(ij) \sum_{klc}            (\sigma_l^{c})^* \langle ck||lj\rangle  \sigma_{ik}^{ab} +
	P(ab) \sum_{lcd}            (\sigma_l^{c})^* \langle bc||dl\rangle  \sigma_{ij}^{ad} +
	P(ij) \sum_{lcd} \frac{1}{2}(\sigma_l^{c})^* \langle ab||id\rangle  \sigma_{jl}^{dc}\nonumber \\ 
	&-&
	P(ab) \sum_{klc} \frac{1}{2}(\sigma_l^{c})^* \langle ak||ij\rangle \sigma_{kl}^{bc} +
	      \sum_{klc} (\sigma_l^{c})^* \langle ck||ji\rangle  \sigma_{kl}^{ab} +
	P(ij) P(ab) \sum_{klc} (\sigma_l^{c})^* \langle bk||li\rangle  \sigma_{jk}^{ca} \nonumber \\
	&-&
	P(ij) P(ab) \sum_{lcd} (\sigma_l^{c})^* \langle ac||dj\rangle \sigma_{il}^{db} -
	            \sum_{lcd} (\sigma_l^{c})^* \langle ab||dl\rangle \sigma_{ij}^{dc} - 
	P(ij)       \sum_{klc} \langle kl||cj\rangle \sigma_k^{c} \sigma_{il}^{ab} \nonumber \\
	&+&
	P(ab) \sum_{kcd}  \langle kb||cd \rangle \sigma_k^{c} \sigma_{ij}^{ad} -
	P(ij) P(ab) \sum_{klc} \langle kl||cj \rangle \sigma_l^{b} \sigma_{ik}^{ac} +
	P(ij) P(ab) \sum_{kcd} \langle kb||cd \rangle \sigma_j^{d} \sigma_{ik}^{ac} \nonumber \\
	&+&
	P(ij) \frac{1}{2} \sum_{klc} \langle kl||ci\rangle \sigma_j^{c} \sigma_{kl}^{ba} -
	P(ab) \frac{1}{2} \sum_{kcd} \langle ka||cd\rangle \sigma_k^{b} \sigma_{ij}^{dc}
	\nonumber \\
	&+&
	P(ab)       \sum_{kl} \frac{1}{2} \langle kl||ij \rangle \sigma_k^a \sigma_l^b -
	P(ij) P(ab) \sum_{kc} \langle ak||cj \rangle \sigma_i^c \sigma_k^b +
	P(ij)       \sum_{cd} \frac{1}{2} \langle ab||cd \rangle \sigma_i^c \sigma_j^d  \nonumber \\
	&-&
	P(ab) \sum_{kc} \frac{1}{3} (\sigma_k^{c})^* \langle ac||ij\rangle \sigma_k^b -
	P(ij) \sum_{kc} \frac{1}{3} (\sigma_k^{c})^* \langle ab||ik\rangle \sigma_j^c.
	\label{hbarabij2}
\end{eqnarray}

The qUCCSD scheme truncates $\bar{H}_{ij}$, $\bar{H}_{ab}$, and
$\bar{H}_{ia,bj}$ in the singles-singles block 
of the excited-state eigenvalue equations to up to the double commutators.
The expressions for $\bar{H}^{\text{qUCCSD}}_{ij}$ thus is given by
\begin{eqnarray}
\bar{H}^{\text{qUCCSD}}_{ij}&=&\bar{H}_{ij}^{0}+\bar{H}_{ij}^{1}+\bar{H}_{ij}^{2}, \\
	\bar{H}_{ij}^{0}&=& f_{ij}, \label{hbarij0}\\
	\bar{H}_{ij}^{1}&=&
	\frac{1}{4}\sum_{kab}\langle ik||ab\rangle \sigma_{jk}^{ab} + 
	\sum_{ka}\langle ik||ja\rangle \sigma_k^{a}+ h.c.,  \label{hbarij1} \\
	\bar{H}_{ij}^{2} &=&
	\left(\sum_{klabc}\frac{1}{2}(\sigma_{kl}^{bc})^* \langle ic||al \rangle \sigma_{jk}^{ab}
	     +\sum_{klmab}\frac{1}{8} (\sigma_{kl}^{ab})^*\langle im||kl \rangle \sigma_{jm}^{ab}
		 + h.c. \right)
	     -\sum_{klmab}\frac{1}{2}(\sigma_{kl}^{ab})^* \langle im||jl \rangle \sigma_{km}^{ab}
	\nonumber \\
	&+&  
	\sum_{klabc}\frac{1}{2}(\sigma_{kl}^{ac})^* \langle ic||jb \rangle \sigma_{kl}^{ab} \nonumber \\
    &+&
    \left(\sum_{kabc}\frac{1}{4} (\sigma_k^b)^* \langle ib||ac \rangle \sigma_{jk}^{ac}
         -\sum_{klab}\frac{1}{2} (\sigma_k^b)^* \langle il||ak \rangle \sigma_{jl}^{ab}
         +\sum_{klab}\frac{1}{2} (\sigma_l^b)^* \langle ik||ja\rangle \sigma_{kl}^{ab}
		 + h.c. \right)
	\nonumber \\
    &+&
	\left(\sum_{kab}\frac{5}{12}\langle ik||ab \rangle \sigma_{j}^{a} \sigma_{k}^{b}
	+\sum_{kab}\frac{1}{2}(\sigma_{k}^{b})^* \langle ib||ak\rangle \sigma_{j}^{a} + h.c. \right)
    -\sum_{kla}(\sigma_{l}^{a})^* \langle ik||jl \rangle \sigma_{k}^{a}
	\nonumber \\
	&+&
    \sum_{kab}(\sigma_{k}^{a})^* \langle ia||jb \rangle \sigma_{k}^{b} .
	\label{hbarij2}
\end{eqnarray}
Similarly, $\bar{H}^{\text{qUCCSD}}_{ab}$ and $\bar{H}^{\text{qUCCSD}}_{ia,bj}$ can be written as
\begin{eqnarray}
\bar{H}^{\text{qUCCSD}}_{ab}&=&\bar{H}_{ab}^{0}+\bar{H}_{ab}^{1}+\bar{H}_{ab}^{2}, \\
	\bar{H}_{ab}^{0}&=& f_{ab}, \label{hbarab0}\\
	\bar{H}_{ab}^{1}&=&
	\left(
	-\sum_{ijc}\frac{1}{4}\langle ij||bc \rangle \sigma_{ij}^{ac}
	+\sum_{ic}\langle ai||bc \rangle \sigma_i^{c} + h.c.\right), \label{hbarab1} \\
	\bar{H}_{ab}^{2}
	&=&
	  \left(
	  -\sum_{ijkcd}\frac{1}{2} (\sigma_{ij}^{cd})^* \langle kd||bj \rangle \sigma_{ik}^{ca}
	  -\sum_{ijcdf}\frac{1}{8} (\sigma_{ij}^{fd})^* \langle df||cb \rangle \sigma_{ij}^{ac}
	  + h.c. \right)
	  +\sum_{ijcdf}\frac{1}{2} (\sigma_{ij}^{fd})^* \langle ad||bc \rangle \sigma_{ij}^{fc} \nonumber \\
    &-&
	  \frac{1}{2}\sum_{ijkcd}(\sigma_{ij}^{cd})^*\langle ka||jb \rangle \sigma_{ik}^{cd}
	\nonumber \\
    &+&
	\left(\sum_{ijkc}\frac{1}{4}(\sigma_j^c)^* \langle ik||bj \rangle \sigma_{ik}^{ac}
	- \sum_{ijcd}\frac{1}{2}(\sigma_j^c)^* \langle ic||bd\rangle \sigma_{ij}^{ad}
	+ \sum_{ijcd}\frac{1}{2}(\sigma_j^d)^* \langle ia||cb\rangle \sigma_{ij}^{cd}
	+ h.c. \right) 
	\nonumber \\
    &+&
	 \left(-\frac{5}{12}\sum_{ijc}\langle ij||bc \rangle \sigma_i^a \sigma_j^c
	 -\sum_{ijc}\frac{1}{2}(\sigma_j^c)^*\langle ic||bj \rangle \sigma_i^a + h.c. \right) 
	 -\sum_{ijc} (\sigma_i^c)^*\langle ja||ib \rangle  \sigma_j^c    \nonumber \\
	&+&
	 \sum_{icd} (\sigma_i^d)^* \langle ad||bc \rangle \sigma_i^c, 
	\label{hbarab2}
\end{eqnarray}
and
\begin{eqnarray}
\bar{H}^{\text{qUCCSD}}_{ia,bj}&=&\bar{H}_{ia,bj}^{0}+\bar{H}_{ia,bj}^{1}+\bar{H}_{ia,bj}^{2}, \\
	\bar{H}_{ia,bj}^{0}&=& \langle ia || bj \rangle, \label{hbariabj0} \\
	\bar{H}_{ia,bj}^{1}&=&
	\frac{1}{2}\sum_{kc}(\sigma_{ik}^{bc})^* \langle ac||jk \rangle +
	 \sum_{c}\langle ai||cb \rangle \sigma_j^{c}
	-\sum_{k}\langle ki||jb \rangle \sigma_k^{a} + h.c.,  \label{hbariajb1} \\
	\bar{H}_{ia,bj}^{2}&=&
	\left(
	       \frac{1}{4}\sum_{klmc}(\sigma_{kl}^{bc})^* \langle im||kl \rangle \sigma_{jm}^{ac}
	      +\frac{1}{4}\sum_{kcde}(\sigma_{ik}^{ce})^* \langle ce||bd \rangle \sigma_{jk}^{ad}
	      -\frac{1}{2}\sum_{klcd}(\sigma_{ik}^{cd})^* \langle lc||kb \rangle \sigma_{jl}^{ad}
	\right.\nonumber \\
	&-&
	       \frac{1}{2}\sum_{klcd}(\sigma_{kl}^{bd})^* \langle id||kc \rangle \sigma_{jl}^{ac}
	      -\frac{1}{4}\sum_{klcd}(\sigma_{kl}^{cd})^* \langle id||bj \rangle \sigma_{kl}^{ca}
	      -\frac{1}{4}\sum_{klcd}(\sigma_{kl}^{cd})^* \langle ia||bl \rangle \sigma_{jk}^{dc}
	\nonumber \\
	&-&\left.         \sum_{klcd}(\sigma_{kl}^{bd})^* \langle ia||kc \rangle \sigma_{jl}^{cd} + h.c. 
	\right) \nonumber \\
	&+&               \sum_{klmc}(\sigma_{lm}^{bc})^* \langle ik||lj \rangle \sigma_{km}^{ac}
          +           \sum_{kcde}(\sigma_{ik}^{de})^* \langle ad||cb \rangle \sigma_{jk}^{ce}
	      +\frac{1}{2}\sum_{klcd}(\sigma_{il}^{cd})^* \langle ka||bl \rangle \sigma_{kj}^{cd}
	\nonumber \\
	&+&    \frac{1}{2}\sum_{klcd}(\sigma_{kl}^{cb})^* \langle ic||dj \rangle \sigma_{kl}^{ad}
	\nonumber \\
	&+& \left(
	      -\frac{1}{2}\sum_{klc}(\sigma_{l}^{c})^* \langle ik||bj \rangle \sigma_{kl}^{ac}
	      +\frac{1}{2}\sum_{kcd}(\sigma_{k}^{d})^* \langle ia||bc \rangle \sigma_{jk}^{cd}
	      -\frac{1}{2}\sum_{klc}(\sigma_{k}^{c})^* \langle il||bk \rangle \sigma_{jl}^{ac}
	\right. \nonumber \\
	&+&
	       \frac{1}{2}\sum_{kcd}(\sigma_{k}^{d})^* \langle id||bc \rangle \sigma_{jk}^{ac}
	      +\frac{1}{2}\sum_{kcd}(\sigma_{i}^{d})^* \langle kd||cb \rangle \sigma_{jk}^{ac}
	      -\frac{1}{2}\sum_{klc}(\sigma_{k}^{b})^* \langle il||kc \rangle \sigma_{jl}^{ac}
	\nonumber \\
	&-& 
	       \frac{1}{4}\sum_{klc}(\sigma_{i}^{c})^* \langle kl||bj \rangle \sigma_{kl}^{ac}
	      +\frac{1}{2}\sum_{klc}(\sigma_{l}^{b})^* \langle ik||cj \rangle \sigma_{kl}^{ac}
	      -\frac{1}{4}\sum_{kcd}(\sigma_{k}^{b})^* \langle ai||cd \rangle \sigma_{jk}^{cd}
	\nonumber \\
	&-&\left.
	       \frac{1}{2}\sum_{kcd}(\sigma_{i}^{d})^* \langle ka||bc \rangle \sigma_{jk}^{cd}
	+ h.c. \right) \nonumber \\
	&+&
	\left(-\frac{2}{3}\sum_{kc}\langle ik||bc \rangle \sigma_{k}^{a}\sigma_{j}^{c}
	      -\frac{1}{2}\sum_{kc}(\sigma_{k}^{c})^* \langle ic||bj \rangle \sigma_{k}^{a} 
	      -\frac{1}{2}\sum_{kc}(\sigma_{k}^{c})^* \langle ia||bk \rangle \sigma_{j}^{c}
	\right. \nonumber \\
	&\red{-}&\left.
	  \red{\sum_{kc}(\sigma_{k}^{b})^* \langle ia || kc \rangle \sigma_{j}^{c}}+h.c. \right)
	      +\sum_{kl}(\sigma_{l}^{b})^* \langle ik||lj \rangle \sigma_{k}^{a}
	      +\sum_{cd}(\sigma_{i}^{d})^* \langle ad||cb \rangle \sigma_{j}^{c},
	\label{hbariabj2}
\end{eqnarray}
respectively. 
$\bar{H}_{ab,ci}$, $\bar{H}_{ia,kj}$, $\bar{H}_{ibc,ajk}$ are involved in the
singles-doubles and doubles-singles block.
They are truncated up to single commutators, i.e.,
$\bar{H}^{\text{qUCCSD}}_{ab,ci}$ is given by
\begin{eqnarray} 
    \bar{H}^{\text{qUCCSD}}_{ab,ci}&=&\bar{H}_{ab,ci}^{0}+\bar{H}_{ab,ci}^{1}, \\
	\bar{H}_{ab,ci}^{0}&=& \langle ab||ci \rangle, \\ 
	\bar{H}_{ab,ci}^{1}&=& P(ab) \sum_{jd}\langle aj||cd\rangle \sigma_{ij}^{bd} 
	+\frac{1}{2} \sum_{kj} \langle jk||ci \rangle
	\sigma_{jk}^{ab} 
	-\frac{1}{2}\sum_{j} (\sigma_j^c)^* \langle ab || ji \rangle \nonumber\\ 
	&+&\sum_{d}\langle ab||cd \rangle \sigma_i^{d} -P(ab) \sum_{j} \langle aj||ci\rangle \sigma_j^{b}. 
	\label{Habci1} 
\end{eqnarray}
$\bar{H}^{\text{qUCCSD}}_{ia,jk}$ can be written as
\begin{eqnarray}
\bar{H}^{\text{qUCCSD}}_{ia,jk}&=&\bar{H}_{ia,jk}^{0}+\bar{H}_{ia,jk}^{1}, \\
	\bar{H}_{ia,jk}^{0} &=& \langle ia || jk \rangle, \\
	\bar{H}_{ia,jk}^{1} &=&
	 P(jk) \sum_{lb}\langle il||jb \rangle \sigma_{kl}^{ab}
	+\frac{1}{2}\sum_{bc}\langle ia||bc \rangle \sigma_{jk}^{bc} 
	+{\frac{1}{2} \sum_b (\sigma_i^b)^* \langle ba || jk \rangle }\nonumber \\
	&-&
	\sum_{l}\langle il||jk \rangle \sigma_l^{a}
	+P(jk) \sum_{b} \langle ai||bj \rangle \sigma_k^{b},
	\label{Hiajk1}
\end{eqnarray}
and the three-body term takes the form
\begin{equation}
	\bar{H}^{\text{qUCCSD}}_{ibc,ajk} = - P(jk) \sum_{l} \langle il||aj \rangle \red{\sigma}_{kl}^{cb} 
	+ P(bc) \sum_{d} \langle ib||ad \rangle \red{\sigma}_{jk}^{dc}.
	\label{hibcajk}
\end{equation}
\red{The contributions from this three-body term to the excited-state eigenvalue
equation is evaluated using the efficient algorithms similar to those within the
EOM-CCSD method,\cite{Stanton1993} i.e., for the evaluation of the contribution
$\sum_{jkbc}\bar{H}_{ibc,ajk}^\mathrm{qUCCSD} C_{jk}^{bc}$ to the singles
residue one first contracts $C_{jk}^{bc}$ with $\sigma_{kl}^{bc}$ or
$\sigma_{jk}^{dc}$ to form one-body intermediates, while for the evaluation of
the contribution $\sum_{jkbc}\bar{H}_{ibc,ajk}^\mathrm{qUCCSD} C_{i}^{a}$
to the doubles residue one first contracts $C_i^a$ with $\langle il || aj
\rangle$ or $\langle ib || ad \rangle$ to form one-body intermediates.}

$\bar{H}_{ij,kl}$ and $\bar{H}_{ab,cd}$ contribute to the doubles-doubles block 
and in the qUCCSD scheme comprise only the bare Hamiltonian integral
\begin{eqnarray}
	\bar{H}^{\text{qUCCSD}}_{ijkl} = \langle ij || kl \rangle,~ 
	\bar{H}^{\text{qUCCSD}}_{ab,cd}= \langle ab || cd \rangle.
\end{eqnarray}
Note that the qUCCSD scheme also uses the bare Hamiltonian integrals
for $\bar{H}_{ij}$, $\bar{H}_{ab}$, and $\bar{H}_{ia,bj}$ in the calculations of the contributions from $\bar{H}_{\mathrm{DD}}$
to the excited-state equations.  $\bar{H}_{iab,jcd}$ and
$\bar{H}_{ija,klb}$ do not contribute to the qUCCSD working equations.

\red{The qUCCSD ground-state amplitude equations are solved using the same
iterative procedure as CCSD, while the excited-state eigencalue equations are
solved using the Davidson algorithms.\cite{Stanton1993,Davidson1975} qUCCSD and CCSD
or EOM-CCSD share ``particle-particle ladder contractions'' of the type
$\sum_{cd} \langle ab || cd \rangle \sigma_{ij}^{cd}$ with a $N_o^2
N_v^4$ scaling and ``ring contraction'' of the type $\sum_{kc} \langle ak ||
ic \rangle \sigma_{jk}^{bc}$ or $\sum_{kc} \langle ik || ac \rangle
\sigma_{jk}^{bc}$ with a $N_o^3N_v^3$ scaling, in which $N_o$ and $N_v$
represent the number of occupied and virtual orbitals, as the most
time-consuming steps. The qUCCSD ground-state amplitude equations involve 
one particle-particle ladder contraction and four ring contractions
per iteration, to be compared with one particle-particle ladder contraction 
and two ring contractions in CCSD. Overall, the computing time of a qUCCSD
ground-state calculation is expected to be around twice that of a CCSD
calculations. The qUCCSD excited-state eigenvalue equations share the same 
particle-particle ladder and ring contractions as EOM-CCSD and thus have 
essentially indentical computational cost per iteration as EOM-CCSD.}


\section{Computational Details}

The qUCCSD method for the calculations of ground-state energies
and excitation energies as detailed in Section II.C have been implemented
in the X2CSOCC module \cite{Liu2018-SOCC} of the CFOUR program\cite{CFOUR,cfour2020} on top of
the previous implementation of the UCC3 method.\cite{Liu2018}
In order to demonstrate the accuracy of the qUCCSD method for 
challenging ground-state problems,
qUCCSD calculations for the equilibrium structures and harmonic frequencies
of \ce{CuH}, \ce{CuF}, and 
\ce{O3} using cc-pVTZ basis sets \cite{Dunning1989,Balabanov2005} have been carried out 
and compared with the corresponding results obtained from CCSD, UCC(4), and UCC3 
calculations. 
The copper-containing molecules have been chosen as examples with
strong orbital-relaxation effects
that have been shown to be difficult to treat using approximate variants of
CC methods.\cite{Bohme1994,Hrusak1997}
The calculations of structural parameters for the ozone molecule,
especially the vibrational frequency for the asymmetric stretching mode
and the ordering of the asymmetric and symmetric stretching frequencies,
played an important role in establishing the CCSD
and CCSD(T) methods.
\cite{Scuseria1989,Stanton1989,Magers1989,Pople1989,Lee1990,Watts1991,Watts1998,Kucharski1999}
In spite of a certain degree of diradical character in ozone, 
CCSD and CCSD(T) can provide qualitatively correct results. It is important for a UCC method with a truncation
of the commutator expansion to have this robustness. 

The classic benchmark set compiled by Trofimov {\it{et al.}} consisting of excitation energies
in \ce{H2O}, \ce{HF}, \ce{N2}, \ce{Ne},
\ce{CH2}, \ce{BH}, and \ce{C2} \cite{Schirmer2002} have been used
to demonstrate the accuracy of qUCCSD excitation energies. 
We have used the same structures
and basis sets
as in the previous calculations
\cite{Koch1995,Christiansen1996,Larsen2001,Hald2001,Schirmer2002}
summarized in the footnotes 81 and 82 of Ref. \cite{Liu2018}
\red{The full configuration interaction (FCI) excitation energies
have been given as reference values. The results obtained
using EOM-CCSD, ADC(3), and UCC3 methods with the same computational scaling
as qUCCSD have also been presented for comparison. We mention
that the CC3 method includes an iterative treatment of triple excitations
and thus is in general more accurate but at the same time more 
time consuming than the qUCCSD method.}
Here \ce{H2O}, \ce{HF}, \ce{N2}, and \ce{Ne}
serve as example molecules for which
the perturbation series converge smoothly,
and \ce{CH2}, \ce{BH}, and \ce{C2} as examples
in the absence of a smooth convergence of the ADC series.  
We focus our discussion on the improvement of the performance
of qUCCSD over the previous UCC3 method. 


\red{Although general characterization of same-symmetry conical
intersections using qUCCSD will have to wait for the implementation of 
analytic gradients and derivative coupling, it is worthwhile mentioning
that the hermitian nature of qUCCSD enables the description of degeneracies
between electronic states. As an example, we have enclosed in the Supporting
Information a qUCCSD calculation of potential energy surfaces in the
immediate vicinity of one of the conical intersection point between the 
$2^1A_1$ and $3^1A_1$ states of the \ce{HOF} molecule. The qUCCSD calculations
show the correct degeneracy at the intersecting point and the correct
linear behavior of the electronic energies with respect to the displacements.
In contrast, EOM-CCSD calculations produce complex eigenvalues when the 
energies of these two state are within 0.03 eV of each other.
}

\section{Results and Discussions}

\subsection{Equilibrium structures and harmonic frequencies
for \texorpdfstring{\ce{CuH}}{}, \texorpdfstring{\ce{CuF}}{}, and \texorpdfstring{\ce{O3}}{}}

Copper-containing molecules serve as excellent avenues to test the robustness
for approximate many-body methods. 
They exhibit significant orbital-relaxation effects, e.g., the largest CCSD singles amplitudes in
CuH and CuF amount to around 0.06. On the other hand, the wavefunctions
are dominated by a single determinant and the CCSD and CCSD(T) methods 
can provide accurate results for properties of CuH and CuF, e.g.,
as shown in Refs. \cite{Bohme1994,Hrusak1997,Cheng2012}
Here we focus our discussion on
the assessment of the qUCCSD, UCC3, and UCC(4) results using the CCSD results
as the reference values. 
As shown in Table \ref{gs}, the UCC3 and UCC(4) results exhibit large
discrepancies compared with the CCSD ones, e.g., the UCC3 harmonic frequency of 739 cm$^{-1}$ for
CuF is more than 100 cm$^{-1}$ greater than the CCSD value of 609 cm$^{-1}$.
In contrast, the qUCCSD results agree closely with the CCSD values, with the deviations in frequencies amounting to 8 cm$^{-1}$ for CuH and 2 cm$^{-1}$ for CuF. 
Interestingly, the UCC3 and UCC(4) calculations of CuH and CuF produced
singles amplitudes larger than 0.2. 
This might be attributed to that UCC3 and UCC(4) have only linear terms involving single excitations
in the amplitude equations, which results in larger $t_1$ amplitudes when attempting to account
for the large orbital-relaxation effects. 
It thus is essential to include the quadratic terms involving single excitations
in the amplitude equations to obtain robust performance.


Ozone is a classic molecule for testing the accuracy of electronic-structure methods.
In particular, the asymmetric stretching frequencies, $\omega_3$, of
\ce{O3} is very sensitive to the treatment of electron correlation.
For example, calculations of $\omega_3$ demonstrated the importance of the fifth-order contribution
in the noniterative triples correction of the CCSD(T) method. \cite{Pople1989}
Although the ground state of ozone possesses certain degree of biradical character,
i.e., the largest $t_2$ amplitude amount to around 0.2, 
the CCSD and CCSD(T) methods can provide quite accurate 
equilibrium structures and vibrational frequencies. \cite{Scuseria1989,Stanton1989,Magers1989,Pople1989,Lee1990,Watts1991,Watts1998,Kucharski1999}
As shown in Table \ref{gs}, the UCC3 and UCC(4) calculations
provide inaccurate results for the structures and harmonic frequencies
of ozone. UCC3 grossly overestimated
$\omega_3$ and
UCC(4) produced an imaginary harmonic frequency for this mode. 
The qUCCSD method obtained structures and vibrational frequencies
in close agreement with the CCSD results, demonstrating the robustness
of the commutator truncation scheme.
As expected, the qUCCSD results is slightly worse than the CCSD ones,
with the latter obtaining the correct ordering of $\omega_2$ and $\omega_3$. \cite{Scuseria1989}
The inclusion of higher commutators is expected to further improve the performance over qUCCSD.

\subsection{Excitation energies of \texorpdfstring{\ce{H2O}}{},
\texorpdfstring{\ce{HF}}{}, \texorpdfstring{\ce{N2}}{}, and \texorpdfstring{\ce{Ne}}{}}

We use \ce{H2O}, \ce{HF}, \ce{N2}, and \ce{Ne} as examples
for which the M{\o}ller-Plesset perturbation series converge smoothly. The excitation
energies for these molecules 
computed using the qUCCSD method
are summarized in Tables \ref{t1}-\ref{t4} together with the corresponding FCI, ADC(3), UCC3,
and EOM-CCSD values. Here we use the FCI values as the reference
and give the other results as the deviation from the FCI values. 
The balanced inclusion of high-order terms in
the qUCCSD scheme provides uniformly better
excitation energies than UCC3. 
The mean absolute deviations of the qUCCSD results amount
to \red{0.12} eV for \ce{H2O}, \red{0.13} eV for \ce{HF}, \red{0.19} eV for \ce{N2}, and
\red{0.18} eV for Ne, which exhibit consistent improvement compared with
the UCC3 values of 0.16 eV for \ce{H2O}, 0.19 eV for \ce{HF}, 0.21 eV for \ce{N2},
and 0.22 eV for Ne. 
The performance of qUCCSD for these molecules is similar to that of EOM-CCSD. 
The absolute mean deviations of the qUCCSD results
with respect to FCI values are slightly larger than those of EOM-CCSD for
\ce{H2O} (by \red{0.04} eV) and \ce{N2} (by \red{0.06} eV) and slightly smaller for
\ce{HF} (by \red{0.03} eV) and \ce{Ne} (by \red{0.03} eV).


\subsection{Excitation energies of
\texorpdfstring{\ce{CH2}}{}, \texorpdfstring{\ce{BH}}{}, and
\texorpdfstring{\ce{C2}}{}}

The computed vertical excitation energies for \ce{CH2} and \ce{BH}
are summarized in Table \ref{t5}-\ref{t6} as examples of simple molecules
for which the ADC series do not converge smoothly. \cite{Schirmer2002} Here the mean and maximum absolute deviations of 
the ADC(3) method with respect to the FCI values
are much larger than 
for the molecules in the previous subsection. 
UCC3 provides better results perhaps because of the iterative solutions
of the ground-state amplitude equations. \cite{Liu2018}
\red{The performance of qUCCSD is similar to that of UCC3 for \ce{BH} and
\ce{CH2}.}
The mean absolute deviation of the qUCCSD 
excitation energies with respect to the FCI results amount to \red{0.07} eV for
\ce{CH2} and \red{0.11} eV for \ce{BH}, to be compared with 0.07 eV
and 0.12 eV in the case of UCC3. 
The mean absolute deviations of qUCCSD
are still greater than those of
EOM-CCSD, by \red{0.05} eV for \ce{CH2} and by \red{0.06} eV for \ce{BH}. 


The ground state of the \ce{C2} molecule has a certain degree of biradical character with the largest $t_2$ amplitude amounting to more than 0.2.
The calculations of excitation energies for \ce{C2} thus serves as
a challenging test for the present truncated UCC-based polarization
propagator methods. 
As shown in Table \ref{t7}, the absolute deviation
of the qUCCSD vertical excitation energies
with respect to the FCI values amount to 
\red{0.38} eV for the $^1\Pi_u$ state, \red{0.53} eV for the $^1\Sigma_u^+$
state, \red{0.54} eV for the a$^3\Pi_u$ state
and \red{0.65} eV for the c$^3\Sigma_u^+$ state.
These are significantly more accurate than the UCC3 values with errors
as large as 0.64 eV, 1.02 eV, 0.74 eV, and 0.88 eV for $^1\Pi_u$, $^1\Sigma_u^+$, a$^3\Pi_u$,
and c$^3\Sigma_u^+$, respectively.
As expected, the qUCCSD method is still not as accurate as the EOM-CCSD method
for the excitation energies for \ce{C2}.
On the other hand, the significant improvement of qUCCSD over UCC3 indicates that
the commutator truncation scheme offers a promising pathway to obtain
robust practical UCC-based methods;
the inclusion of triple and higher commutators is expected to further 
improve the accuracy of the method. 

\section{Summary and Outlook} 

We develop a self-consistent polarization propagator method using a quadratic unitary coupled-cluster singles and doubles (qUCCSD) parameterization for the ground state wavefunction and the excitation manifold. 
Benchmark calculations of ground-state properties and excitation energies for representative small molecules show that the qUCCSD scheme using a commutator truncation scheme exhibits a uniform improvement
of the accuracy and robustness over the previous UCC3 method derived using M{\o}ller-Plesset perturbation theory. 
The future work will be focused on an implementation of \red{the qUCCSD scheme
and its analytic gradients and derivative coupling within tensor contraction
engines well developed for the non-relativistic CC machinery} to enable
extensive molecular applications and the development of a cubic UCCSD (cUCCSD) scheme, i.e., the UCCSD[3$\mid$3,2,1] scheme, to further improve the accuracy and robustness. 

\section{Acknowledgement}
 
This work has been supported by the Department of Energy, Office of Science,
Office of Basic Energy Sciences under Award Number DE-SC0020317. All
computations in this work were carried out using Maryland Advanced Research
Computing Center (MARCC). L. C. is greatly indebted to Debashis Mukherjee
(Kolkata), J{\"u}rgen Gauss (Mainz), and John Stanton (Gainesville) for
stimulating discussions and support. 
L. C. is grateful to Gustavo Scuseria (Houston) for helpful discussions about the basis-set effects in calculations for vibrational frequencies of ozone. 

\section{Data Availability Statement}

The data that supports the findings of this study are available within the article.


\clearpage
\begin{center}
   \begin{threeparttable}
	\caption{Computed equilibrium bond lengths (in \AA), bond
	angle (in degree), and harmonic frequencies (in cm$^{-1}$) of \ce{CuH},
	\ce{CuF}, and \ce{O3}. The cc-pVTZ basis sets were used for all the
	calculations presented here. The $1s$ electrons of O and $1s$, $2s$, $2p$, $3s$, $3p$ electrons of Cu
	have been kept frozen in the electron-correlation calculations.}
	\label{gs} 
    \tabcolsep=6pt
	\begin{tabular}[t]{@{}ccccccccccccc@{}}
	\hline\hline  
	\multicolumn{2}{c}{\multirow{2}{*}{method}} 
	&\multicolumn{2}{c}{\ce{CuH}} & &\multicolumn{2}{c}{\ce{CuF}} & 
    &\multicolumn{5}{c}{\ce{O3}} \\
     \cline{3-4} \cline{6-7} \cline{9-13}
	& &$R_{\ce{Cu-H}}$ &$\omega_e$ & &$R_{\ce{Cu-F}}$  &$\omega_e$ & 
	  &$R_{\ce{O-O}}$ &$\theta$ &$\omega_{1,e} (a_1)$ &$\omega_{2,e} (a_1)$ &$\omega_{3,e} (b_2)$\\
	\hline 
	UCC(4)                   & &1.4616 &2052 & &1.6998 &646 & &1.3142 &117.1 &560 &876  &1922$i$ \\
	UCC3                     & &1.4877 &1948 & &1.7367 &739 & &1.2659 &117.9 &674 &1033 &4698    \\
	qUCCSD                   & &1.4891 &1829 & &1.7686 &607 & &1.2488 &117.5 &767 &1279 &1314    \\
	CCSD                     & &1.4888 &1837 & &1.7669 &609 & &1.2499 &117.6 &763 &1278 &1266    \\
	\hline\hline 
    \end{tabular}
   \end{threeparttable}
\end{center}

\begin{center}
   \begin{threeparttable}
	\caption{Computed vertical excitation energies (in eV) of the \ce{H2O} molecule. 
	The UCC3, qUCCSD, 
	ADC(3), and CCSD values are presented as
	the differences relative to the corresponding FCI values.
	$\bar{\Delta}_\mathrm{abs}$
	and $\Delta_\mathrm{max}$ denote the mean absolute
	error and maximum absolute error relative to the FCI results, respectively.
    \red{The $1s$ electrons of O have been kept frozen in the
	electron-correlation calculations.}}
	\label{t1} 
	\tabcolsep=20pt
	\begin{tabular}[t]{@{}lrrrrr@{}}
	\hline\hline 
		State & FCI\tnote{a}  &ADC(3)\tnote{b} &UCC3 &qUCCSD &CCSD\tnote{a} \\
	\hline
	2\ $^1A_1$  & 9.87  &0.14  &0.20  &\red{0.13}  &-0.07\\
    1\ $^1B_1$  & 7.45  &0.13  &0.18  &\red{0.14}  &-0.07\\
    1\ $^1B_2$  &11.61  &0.18  &0.23  &\red{0.15}  &-0.09\\
    1\ $^1A_2$  & 9.21  &0.17  &0.20  &\red{0.15}  &-0.09\\
    1\ $^3B_1$  & 7.06  &0.09  &0.14  &\red{0.11}  &-0.08\\
    1\ $^3A_2$  & 9.04  &0.14  &0.19  &\red{0.14}  &-0.08\\
    1\ $^3A_1$  & 9.44  &0.10  &0.15  &\red{0.10}  &-0.08\\
    2\ $^3A_1$  &10.83  &0.01  &0.04  &\red{0.06}  &-0.11\\
    2\ $^3B_1$  &11.05  &0.11  &0.14  &\red{0.11}  &-0.09\\
    1\ $^3B_2$  &11.32  &0.13  &0.17  &\red{0.11}  &-0.08\\
	$\bar{\Delta}_{\mathrm{abs}}$ & -- &0.12 &0.16 &\red{0.12} &0.08 \\
	$\Delta_{\mathrm{max}}$       & -- &0.18 &0.23 &\red{0.15} &0.11 \\
	\hline\hline
    \end{tabular}
    \begin{tablenotes}
	  \item[a] Results for the singlet states are from Ref.\citenum{Christiansen1996}
		  and those for the triplet states are from Ref. \citenum{Larsen2001}.
	  \item[b] Ref.\citenum{Schirmer2002}
	\end{tablenotes}
   \end{threeparttable}
\end{center}

\begin{center}
   \begin{threeparttable}
	   \caption{Computed vertical excitation
	energies (in eV) of the \ce{HF} molecule. 
	The difference of ADC(3), UCC(3), qUCCSD, and CCSD results 
	relative to the corresponding FCI values are presented. 
	$\bar{\Delta}_\mathrm{abs}$
	and $\Delta_\mathrm{max}$ denote the mean absolute
	error and maximum absolute error relative to the FCI results, respectively.
    \red{The $1s$ electrons of F have been kept frozen in the
	electron-correlation calculations.}}
	\label{t2} 
	\tabcolsep=20pt
	\begin{tabular}[t]{@{}lrrrrr@{}}
	\hline\hline 
	State &FCI\tnote{a}  &ADC(3)\tnote{b} &UCC3 &qUCCSD &CCSD\tnote{a} \\
	\hline
	1\ $^1\Pi$      &10.44  &0.18  &0.23  &\red{0.15}  &-0.14\\
    2\ $^1\Pi$      &14.21  &0.19  &0.23  &\red{0.16}  &-0.15\\
    2\ $^1\Sigma^+$ &14.58  &0.10  &0.17  &\red{0.07}  &-0.11\\
    1\ $^1\Delta  $ &15.20  &0.12  &0.16  &\red{0.12}  &-0.17\\
    1\ $^1\Sigma^-$ &15.28  &0.12  &0.15  &\red{0.12}  &-0.18\\
    3\ $^1\Pi$      &15.77  &0.23  &0.25  &\red{0.17}  &-0.18\\
    3\ $^1\Sigma^+$ &16.43  &0.37  &0.36  &\red{0.24}  &-0.14\\
    1\ $^3\Pi$      &10.04  &0.14  &0.20  &\red{0.13}  &-0.15\\
    1\ $^3\Sigma^+$ &13.54  &0.05  &0.09  &\red{0.02}  &-0.13\\
    2\ $^3\Pi$      &14.01  &0.19  &0.23  &\red{0.16}  &-0.16\\
    2\ $^3\Sigma^+$ &14.46  &0.07  &0.11  &\red{0.09}  &-0.21\\
    1\ $^3\Delta$   &14.93  &0.10  &0.13  &\red{0.11}  &-0.19\\
    1\ $^3\Sigma^-$ &15.25  &0.12  &0.16  &\red{0.12}  &-0.18\\
    3\ $^3\Pi$      &15.57  &0.22  &0.25  &\red{0.17}  &-0.19\\
	$\bar{\Delta}_{\mathrm{abs}}$ & -- &0.16 &0.19 &\red{0.13} &0.16 \\
	$\Delta_{\mathrm{max}}$       & -- &0.37 &0.36 &\red{0.24} &0.21 \\
	\hline\hline
    \end{tabular}
        \begin{tablenotes}
	  \item[a] Ref.\citenum{Larsen2001}
	  \item[b] Ref.\citenum{Schirmer2002}
	\end{tablenotes}
   \end{threeparttable}
\end{center}

\begin{center}
   \begin{threeparttable}
   	   \caption{Computed vertical excitation energies (in eV) of the \ce{N$_2$} molecule. 
				The ADC(3), UCC3, qUCCSD, and CCSD values are presented as the differences
				relative to the corresponding FCI values. $\bar{\Delta}_\mathrm{abs}$
				and $\Delta_\mathrm{max}$ denote the mean absolute error and maximum absolute 
				error relative to the FCI results, respectively.
                \red{The $1s$ electrons of N have been kept frozen in the
				electron-correlation calculations.}}
	\label{t3} 
	\tabcolsep=20pt
	\begin{tabular}[t]{@{}lrrrrr@{}}
	\hline\hline 
	State &FCI\tnote{a}  &ADC(3)\tnote{b} &UCC3 &qUCCSD &CCSD\tnote{a} \\
	\hline
	1\ $^1\Pi_g$        & 9.58  &-0.17  &-0.08  &\red{-0.06}  & 0.08\\
    1\ $^1\Sigma_u^-$   &10.33  &-0.33  &-0.27  &\red{-0.23}  & 0.14\\
    1\ $^1\Delta_u$     &10.72  &-0.37  &-0.25  &\red{-0.21}  & 0.18\\
    1\ $^1\Pi_u$        &13.61  &-0.23  &-0.25  &\red{-0.29}  & 0.40\\
    1\ $^3\Sigma_u^+$   & 7.90  &-0.19  &-0.26  &\red{-0.24}  &-0.02\\
    1\ $^3\Pi_g$        & 8.16  &-0.29  &-0.13  &\red{-0.11}  & 0.06\\
    1\ $^3\Delta_u$     & 9.19  &-0.27  &-0.27  &\red{-0.24}  & 0.07\\
    1\ $^3\Sigma_u^-$   &10.00  &-0.29  &-0.25  &\red{-0.22}  & 0.19\\
    1\ $^3\Pi_u$        &11.44  &-0.19  &-0.11  &\red{-0.12}  & 0.10\\
	$\bar{\Delta}_{\mathrm{abs}}$ & -- &0.26 &0.21 &\red{0.19}  & 0.13\\
	$\Delta_{\mathrm{max}}$       & -- &0.37 &0.27 &\red{0.29}  & 0.40\\
	\hline\hline
    \end{tabular}
            \begin{tablenotes}
	  \item[a] Results for the singlet states are from Ref.\citenum{Christiansen1996}
		  and those for the triplet states are from Ref. \citenum{Larsen2001}.
	  \item[b] Ref.\citenum{Schirmer2002}
	\end{tablenotes}
   \end{threeparttable}
\end{center}
\begin{center}
	 \begin{threeparttable}
	 \caption{Computed vertical excitation energies (in eV) of the \ce{Ne} atom. 
        The ADC(3), UCC3, qUCCSD, and CCSD values are presented as the
		differences relative to the corresponding FCI values.  $\bar{\Delta}_\mathrm{abs}$
		and $\Delta_\mathrm{max}$ denote the mean absolute error and maximum
		absolute error relative to the FCI results, respectively. 
        \red{The $1s$ electrons of Ne have been kept frozen in the
		electron-correlation calculations.}}
		\label{t4} 
	\tabcolsep=20pt
	\begin{tabular}[t]{@{}lrrrrr@{}}
	\hline\hline 
	State &FCI\tnote{a}  &ADC(3)\tnote{b} &UCC3  &qUCCSD  &CCSD\tnote{a} \\
	\hline
	1\ $^1P$  &16.40 &0.17 &0.16 &\red{0.14} &-0.24 \\
    1\ $^1D$  &18.21 &0.18 &0.18 &\red{0.15} &-0.25 \\
    2\ $^1P$  &18.26 &0.18 &0.17 &\red{0.15} &-0.25 \\
    2\ $^1S$  &18.48 &0.27 &0.27 &\red{0.20} &-0.24 \\
    3\ $^1S$  &44.05 &0.35 &0.35 &\red{0.32} &-0.17 \\
    1\ $^3P$  &18.70 &0.13 &0.16 &\red{0.11} &-0.24 \\
    1\ $^3S$  &19.96 &0.10 &0.13 &\red{0.10} &-0.26 \\
    1\ $^3D$  &20.62 &0.13 &0.17 &\red{0.12} &-0.23 \\
    2\ $^3P$  &20.97 &0.13 &0.17 &\red{0.12} &-0.24 \\
    2\ $^3S$  &45.43 &0.40 &0.44 &\red{0.36} &-0.10 \\
	$\bar{\Delta}_{\mathrm{abs}}$ & -- &0.20 &0.22 &\red{0.18} &0.22 \\
	$\Delta_{\mathrm{max}}$       & -- &0.40 &0.44 &\red{0.36} &0.25 \\
	\hline\hline
    \end{tabular}
                \begin{tablenotes}
	  \item[a] Results for the singlet states are from Ref.\citenum{Koch1995}
		  and those for the triplet states are from Ref. \citenum{Larsen2001}.
	  \item[b] Ref.\citenum{Schirmer2002}
	\end{tablenotes}
   \end{threeparttable}
\end{center}
\begin{center}
	   \begin{threeparttable}
	    \caption{Computed vertical excitation energies (in eV) of the \ce{CH2} molecule. 
		The ADC(3), UCC3, qUCCSD, and CCSD values are presented as the differences relative to
		the corresponding FCI values.  $\bar{\Delta}_\mathrm{abs}$
		and $\Delta_\mathrm{max}$ denote the mean absolute
		error and maximum absolute error relative to the FCI results, respectively.}
	\label{t5} 
	\tabcolsep=20pt
	\begin{tabular}[t]{@{}lrrrrrr@{}}
	\hline\hline 
	State &FCI\tnote{a} &ADC(3)\tnote{b} &UCC3 &qUCCSD &CCSD\tnote{a} \\
	\hline
	3\ $^1A_1$        & 6.51  &-0.31  &-0.05  &\red{-0.06}  &-0.01\\
	4\ $^1A_1$        & 8.48  &-0.29  &-0.04  &\red{-0.04}  &-0.02\\
    1\ $^1B_2$        & 7.70  &-0.24  & 0.01  &\red{ 0.01}  & 0.01\\
    1\ $^1B_1$        & 1.79  &-0.55  &-0.10  &\red{-0.11}  &-0.01\\
    1\ $^1A_2$        & 5.85  &-0.42  &-0.09  &\red{-0.08}  & 0.01\\
    1\ $^3A_1$        & 6.39  &-0.31  &-0.06  &\red{-0.06}  &-0.01\\
    2\ $^3A_1$        & 8.23  &-0.38  &-0.09  &\red{-0.08}  &-0.03\\
    3\ $^3A_1$        & 9.84  &-0.31  &-0.07  &\red{-0.08}  & 0.01\\
    2\ $^3B_2$        & 7.70  &-0.31  &-0.06  &\red{-0.06}  &-0.06\\
    1\ $^3B_1$        &-0.01  &-0.61  &-0.14  &\red{-0.13}  &-0.03\\
	2\ $^3B_1$        & 8.38  &-0.41  &-0.02  &\red{-0.02}  & 0.01\\
    1\ $^3A_2$        & 4.79  &-0.44  &-0.10  &\red{-0.10}  & 0.00\\
	$\bar{\Delta}_{\mathrm{abs}}$ &   --   &0.38  &0.07  &\red{0.07} &0.02 \\
	$\Delta_{\mathrm{max}}$       &   --   &0.61  &0.14  &\red{0.13} &0.06 \\
	\hline\hline
    \end{tabular}
                        \begin{tablenotes}
	  \item[a] Results for the singlet states are from Ref.\citenum{Koch1995}
		  and those for the triplet states are from Ref. \citenum{Hald2001}.
	  \item[b] Ref.\citenum{Schirmer2002}
	\end{tablenotes}
   \end{threeparttable}
\end{center}
\begin{center}
   \begin{threeparttable}
	   	\caption{Computed vertical excitation energies (in eV) of the \ce{BH} molecule. 
		The ADC(3), UCC3, qUCCSD, and CCSD values are presented as the
		differences relative to the corresponding FCI values.
		$\bar{\Delta}_\mathrm{abs}$ and $\Delta_\mathrm{max}$ denote the mean absolute
		error and maximum absolute error relative to the FCI results, respectively.}
	\label{t6} 
	\tabcolsep=20pt
	\begin{tabular}[t]{@{}lrrrrr@{}}
	\hline\hline 
	State &FCI\tnote{a}  &ADC(3)\tnote{b}  &UCC3 &qUCCSD &CCSD\tnote{a} \\
	\hline
	1\ $^1\Pi$               &2.94  &-0.61  &-0.10  &\red{-0.10} & 0.02\\
    2\ $^1\Sigma^+$          &6.38  &-0.43  &-0.07  &\red{-0.07} & 0.04\\
    2\ $^1\Pi$               &7.47  &-0.51  &-0.14  &\red{-0.11} & 0.04\\
    4\ $^1\Sigma^+$          &7.56  &-0.54  &-0.16  &\red{-0.13} & 0.19\\
	3\ $^1\Pi$               &8.24  &-0.50  &-0.15  &\red{-0.12} & 0.04\\
    1\ $^3\Pi$               &1.31  &-0.62  &-0.11  &\red{-0.10} &-0.01\\
    1\ $^3\Sigma^+$          &6.26  &-0.47  &-0.10  &\red{-0.10} & 0.03\\
    2\ $^3\Sigma^+$          &7.20  &-0.49  &-0.10  &\red{-0.09} & 0.02\\
    2\ $^3\Pi$               &7.43  &-0.51  &-0.14  &\red{-0.13} & 0.00\\
    3\ $^3\Sigma^+$          &7.62  &-0.52  &-0.14  &\red{-0.13} & 0.05\\
    3\ $^3\Pi$               &7.92  &-0.45  &-0.12  &\red{-0.15} & 0.08\\
	$\bar{\Delta}_{\mathrm{abs}}$ & --  &0.51 &0.12 &\red{0.11}  &0.05\\
	$\Delta_{\mathrm{max}}$       & --  &0.62 &0.16 &\red{0.15}  &0.19\\
	\hline\hline
    \end{tabular}
    \begin{tablenotes}
	  \item[a] Results for the singlet states are from Ref.\citenum{Koch1995}
		  and those for the triplet states are from Ref. \citenum{Larsen2001}.
	  \item[b] Ref.\citenum{Schirmer2002}
	\end{tablenotes}
   \end{threeparttable}
\end{center}

\begin{center}
   \begin{threeparttable}
	\caption{Computed vertical excitation energies (in eV) of the \ce{C2} molecule. 
	The UCC3, qUCCSD, and CCSD results are given as the difference relative to 
	the corresponding FCI values. 
    \red{The $1s$ electrons of C have been kept frozen in the electron-correlation calculation.}}
	\label{t7} 
	\tabcolsep=20pt
	\begin{tabular}[t]{@{}lrrrr@{}}
	\hline\hline 
    State &FCI\tnote{a}  &UCC3 &qUCCSD &CCSD\tnote{a}\\
	\hline
	$^1\Pi_u$      &1.39	&-0.64	&\red{-0.38}  & 0.09\\
    $^1\Sigma_u^+$ &5.60	&-1.02	&\red{-0.53}  & 0.20\\
   a$^3\Pi_u$      &0.31	&-0.74	&\red{-0.54}  &-0.03\\
   c$^3\Sigma_u^+$ &1.21	&-0.88	&\red{-0.65}  &-0.44\\
	\hline\hline
    \end{tabular}
        \begin{tablenotes}
	  \item[a] Results for the singlet states are from Ref.\citenum{Christiansen1996}
		  and those for the triplet states are from Ref. \citenum{Larsen2001}.
	\end{tablenotes}
   \end{threeparttable}
\end{center}

\clearpage
\bibliography{qUCCSD}
\end{document}